\documentclass[namedreferences]{solarphysics}
\usepackage[optionalrh,solaromanenum]{spr-sola-addons} 
\usepackage{graphicx}                    
\usepackage{color}                       
\usepackage{url}                         

\newcommand{\aap}{    {\it Astron. Astrophys.}}

\newcommand{\apj}{    {\it Astrophys. J.}}
\newcommand{\apjl}{   {\it Astrophys. J. Lett.}}

\newcommand{\solphys}{{\it Solar Phys.}}

\begin{document}

\begin{article}

\begin{opening}

\title{On Dynamics of G-band Bright Points}

%
\author{M.~\surname{Bodn\'arov\'a}$^{1}$\sep
        D.~\surname{Utz}$^{2,3}$\sep
        J.~\surname{Ryb\'ak}$^{1}$      
       }

%
\runningauthor{Bodn\'arov\'a et al.}
\runningtitle{On Dynamics of G-band Bright Points}

%
 \institute{$^{1}$ Astronomical Institute, Slovak Academy of Sciences, Tatransk\'a Lomnica, SK 059 60, Slovak Republic
                     \newline email: \url{mbodnarova@astro.sk} \\ 
            $^{2}$ Instituto de Astrof\'{i}sica de Andaluc\'{i}a (CSIC), 
Apdo. de Correos 3004, E--18080 Granada, Spain
                     \newline email: \url{utz@iaa.es} \\
            $^{3}$ IGAM/Institute of Physics, University of Graz, 
Universit{\"a}tsplatz 5, A--8010 Graz, Austria
                    \newline email: \url{dominik.utz@uni-graz.at}\\
             }

\begin{abstract}
Various parameters describing dynamics of G-band bright points (GBPs) were derived from 
G-band images, acquired by the \textit{Dutch Open Telescope} (DOT), of a quiet region close to the disk
center. Our study is based on four commonly used diagnostics (an effective velocity, a change in the effective velocity, a change in
the direction angle and a centrifugal acceleration) and two new ones (a rate of motion and a time lag between
recurrence of GBPs). The results concerning the commonly
used parameters are in agreement with previous studies for a comparable spatial and temporal resolution
of the used data. The most probable value of the effective velocity is $\sim$0.9\,km\,s$^{-1}$, whereas we found a
deviation of the effective velocity distribution from the expected Rayleigh function for velocities in the
range from 2 to 4\,km\,s$^{-1}$. The change in the effective velocity distribution is consistent with a Gaussian
one with FWHM = 0.079\,km\,s$^{-2}$. The distribution of the centrifugal acceleration exhibits a highly exponential nature (a symmetric
Gaussian centered at the zero value). To broaden our understanding of
dynamics of GBPs, two new parameters were defined: the real displacement between their appearance and disappearance (a rate of motion) and the frequency of their recurrence at the same locations (a time lag). For $\sim$45\% of the
tracked GBPs, their displacement was found to be small compared to their size (the rate of motion smaller than one). The locations of the tracked GBPs mainly cover the boundaries of supergranules representing the network, and there is no significant difference in the locations of GBPs with small ($m < 1$) and large ($m > 2$) values of the rate of motion.
We observed a difference in the overall trend of the obtained distribution for the values of the time lag smaller (slope of the trend line being -0.14) and greater (-0.03) than $\sim$7\,min. The time lags mostly lie within the interval of $\sim$2-3\,min, with those up to $\sim$4\,min being more abundant than longer ones. Results for both new parameters indicate that the locations of different dynamical types of GBPs (stable/farther traveling or with short/long lifetimes) are bound to the locations of more stable and long-living magnetic field concentrations. Thus, the disappearance/reappearance of the tracked GBPs cannot be perceived as the disappearance/reappearance of their corresponding magnetic field concentrations.

\end{abstract}

%
\keywords{Magnetic fields, Photosphere; Supergranulation}

\end{opening}

%

\section{Introduction}
     \label{S-Introduction}
\par
High-resolution G-band images of the photosphere exhibit small bright features whose diameter is
less than 300 km \cite{1995ApJ...454..531B,2006SoPh..237...13M,2009A&A...498..289U}. These features are called
bright points, or G-band bright points (GBPs). They were for the first time identified in 1980, in filtergrams
taken through a narrow-band interference filter, centered at 4308\,\AA\ \cite{1984SoPh...94...33M}.
GBPs are usually located in the intergranular dark lanes \cite{1987ApJ...317..892T} and
concentrated in active regions, or at the border of the supergranules in the quiet Sun \cite{2004ApJ...609L..91S}. \inlinecite{1996ApJ...463..365B} reported that GBPs are co-spatial with magnetic field
concentrations, being associated with strong magnetic fields of about 1.5 $\times$ $10^{3}$\,G \cite{2007A&A...472..607B,2010ApJ...723..787V}. Several GBP properties (like their lifetimes, sizes and
velocities) were quantified so far, notably in papers of \inlinecite{1994A&A...283..232M}, \inlinecite{1998ApJ...495..973B}, \inlinecite{2003ApJ...587..458N}, \inlinecite{2006SoPh..237...13M}, \citeauthor{2009A&A...498..289U} (\citeyear{2009A&A...498..289U}, \citeyear{2010A&A...511A..39U}). Numerical simulations
\cite{2004ApJ...610L.137C,2005A&A...430..691S,2007A&A...469..731S,2010A&A...509A..76D}
indicate that these small-scale concentrations of magnetic flux originate from the interaction between
magnetic fields and granular convective motions.
\par
It is assumed that rapid GBP foot-point motion can excite MHD waves, which could contribute,
in a significant way, to the heating of the solar corona \cite{1993SoPh..143...49C,1994A&A...283..232M}. In
addition, nanoflares can also be triggered by these flux tube motions
\cite{1988ApJ...330..474P}. Therefore, GBPs, as small-scale features of significant magnetic flux in the photosphere, are of importance to be studied properly in order to understand the solar magnetism and dynamics of
the outer solar atmosphere.
The aim of this paper is to present a focused study of various commonly used as well as a few newly introduced parameters
describing dynamics of GBPs, identified and tracked by an automatic algorithm. We focus on those dynamical
properties of GBPs (such as their effective velocity, centrifugal acceleration, etc.), which could play a crucial role in
an estimation of the energy available for transfer into higher layers of the solar atmosphere in a form of
MHD waves and/or nanoflares.
\par
The paper is organized as follows. Section 2 contains description of the used data recorded by the DOT and
their processing by the algorithm developed by Utz (\opencite{2009A&A...498..289U}, \citeyear{2010A&A...511A..39U}). In Section 3, we summarize our results for various parameters describing dynamics of GBPs. Section 4 presents a comparison of our results with previous results in the field. Finally, Section 5 is reserved for concluding remarks.

\section{Observation, Data Processing and General Statistical Results}
     \label{S-Observation}
\par
We used a data set of speckle-reconstructed images of the quiet solar photosphere taken in the G-band (430\,nm) and recorded 
by the \textit{Dutch Open Telescope} (DOT: \opencite{2004A&A...413.1183R}. The time sequence was collected on 
19 October 2005, at 09:55-11:05 UT, under good seeing conditions from a network region close to the disk
center (location N00/W09, $\mu$ = 0.983). The data set consists of 142 speckle-reconstructed images with
$r_{0}$ $\geq$ 7\,cm and with a cadence of 30 s. Each image was reconstructed from a burst of 100 images, taken
with a rate of 6 images per second \cite{2001ASPC..236..431S}. Speckle-reconstructed images have a field of
view of 79 $\times$ 58\,arcsec, with a spatial sampling of 0.071 arcsec per pixel.
\par
To identify and track GBPs in these speckle-reconstructed G-band images, we used an automated
identification algorithm developed by Utz (\opencite{2009A&A...498..289U}, \citeyear{2010A&A...511A..39U}). The algorithm identifies and tracks
GBPs and consists of three processing steps. The first, segmentation, step is based on the idea of following
contours of the features from their brightest pixels down to their faintest ones in consecutive steps. 
The next, identification, step, takes the brightness gradient of the segments into account. MBPs are peaked features 
in intensity, showing therefore huge brightness gradients, while granules are more flat in intensity and hence posses small brightness gradients. The brightness gradients of all the found segments are calculated and then a 
threshold criterion is applied to identify MBPs and discern them from granules. In the final, time series
generation, step the identified GBPs are tracked in consecutive images and their properties are analyzed.
During the tracking process some of the identified GBPs were excluded, especially from the beginning
and the end of the data set, because they were not observed for their whole existence.
For our study of dynamics of GBPs only single GBPs were taken into account (apart from splitting or merging of GBPs).
\par
As a result of the application of the above-described algorithm on our data set, we obtained a set
of 4017 tracked GBPs consisting of 26238 individual "identifications" of GBPs on 142 speckle-reconstructed G-band images. Each of these 26238 "identifications" is described by a set of parameters,
including its position within the field of view of the DOT, its size and its brightness. Furthermore, the
algorithm allowed us to evaluate some basic statistical properties of the tracked GBPs, namely their
sizes (an average radius of 245 $\pm$ 38\,km), lifetimes (an average lifetime of 3.0 $\pm$ 2.7\,min) and 
velocities (the median of
velocity 1.38\,km\,s$^{-1}$). For more details on the obtained distributions for the above-listed parameters, the interested reader is referred to \inlinecite{2010CEAB...34...25B}.
\par
The most important parameter for our study of dynamics of GBPs, allowing us to define
the locations of the tracked GBPs throughout their existence, was the position of each "identification"
of the GBPs. Such a position was estimated by the algorithm as the
barycenter of brightness computed from all pixels on the speckle-reconstructed G-band image making up the
"identification" of the GBP. Following this procedure, the accuracy of the estimate of the position
of a GBP is greater than that of a simple estimation of a central pixel, or the positional accuracy
given by the brightest pixel. This is due to the fact that the barycenter coordinates can be determined on
a sub-pixel resolution \cite{2010A&A...511A..39U}.
\par
\begin{figure} 
\centerline{\includegraphics[width=\textwidth,clip=]{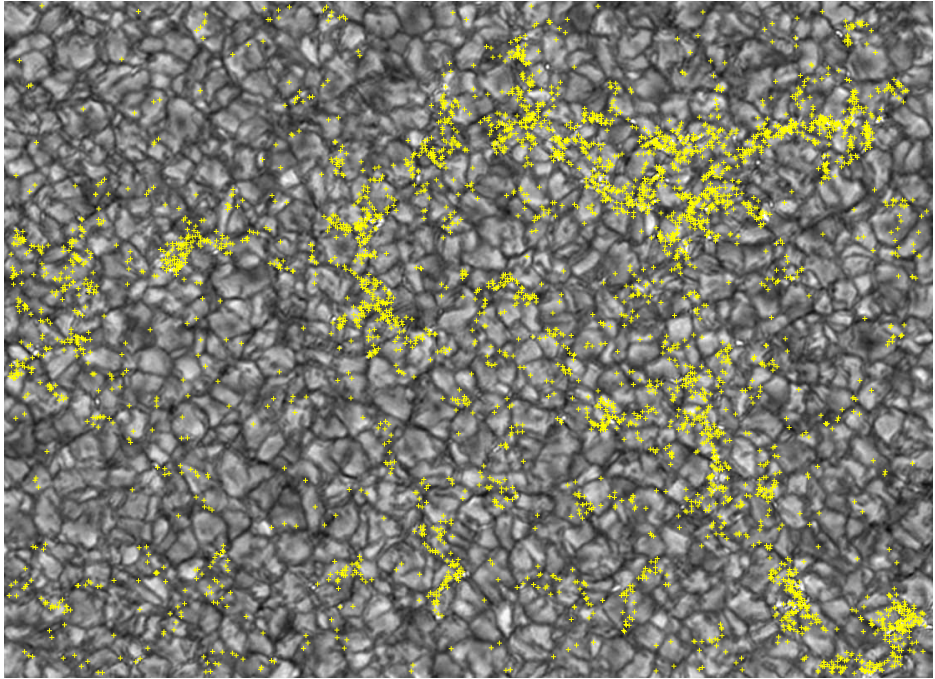}}
\caption{The full field of view of the DOT (79 $\times$ 58 arcsec), with crosses indicating the locations of the
first "identifications" of all tracked GBPs. The field of view of the DOT is illustrated by the second frame
of the G-band data set.}
\label{fig:1}
\end{figure}

As an illustration, Figure \ref{fig:1} shows the obtained locations of the first "identifications" of all tracked
GBPs within the field of view of the DOT. The
observed locations of the tracked GBPs are not distributed homogeneously across the whole field of
view; rather, they seem to prefer grouping in certain regions. Based on the
comparison of G-band images with simultaneous images in Ca II H (397\,nm), obtained by the DOT as well, we
can state that in most cases the areas with the highest density of GBPs coincide with the
outlines of supergranules, \textit{i.e.} the magnetic network (see \textit{e.g} \opencite{2012SoPh..280..407R}).

%

\section{Dynamics of G-band Bright Points}
     \label{S-Dynamics}
\par
We evaluated movement of GBPs based on the statistical analysis of four classical parameters: an effective velocity, a change in the effective velocity (acceleration), a change in direction angle
and a centrifugal acceleration (see \textit{e.g} \opencite{2003ApJ...587..458N}). In addition, we defined two new
parameters: a rate of motion and a time lag between recurrence of GBPs. These parameters can help us to
deepen our understanding of the movement of GBPs and their lasting presence at certain locations.

\subsection{EFFECTIVE VELOCITY} 
  \label{S-Effective velocity}
\par
The most common indicator of the dynamics of GBPs is their effective velocity \begin{math}v=\sqrt{v^{2}_{x} + v^{2}_{y}}\end{math}, where $v_x$ and $v_y$ are the $x$- and $y$-components of the velocity vector. The obtained distributions of both $v_x$ and $v_y$ are of a Gaussian shape and in both cases the Gaussian is symmetrically centered at the zero value (for more details see \opencite{2010CEAB...34...25B}).
\par
For the studied "identifications" of GBPs, the effective velocities were found to lie in the range from 0.0 to 6.0\,
km\,s$^{-1}$. Their distribution is shown in Figure 2. Obviously, most numerous are low velocities in the
range of $\sim$0.5-2.0\,km\,s$^{-1}$, only $\sim$10\% of the measured velocities are greater than 3\,km\,s$^{-1}$ and only 3\% of them surpass 4\,km\,s$^{-1}$. The measured mean value of the effective velocity is 1.62 $\pm$ 1.06\,km\,s$^{-1}$, its median value
is 1.38\,km\,s$^{-1}$ and the most probable value amounts to $\sim$0.9\,km\,s$^{-1}$. This distribution is a non-symmetric one and could be fitted by a Rayleigh distribution (\textit{e.g}
\opencite{2003ApJ...587..458N} or \opencite{2010A&A...511A..39U}) in the form 

\begin{displaymath}
f(v,\sigma)=\frac{v}{\sigma^{2}}exp(\frac{-v^{2}}{2\sigma^{2}}),    
\end{displaymath}

\noindent with $\sigma$ being the standard deviation.
In probability theory and statistics, the Rayleigh distribution is a continuous probability distribution for positive-valued random variables. It is often observed when the overall magnitude of a vector is related to its directional components (\textit{e.g}, the horizontal and vertical coordinates chosen independently from the standard normal distribution). 
\par
\begin{figure} 
\centerline{\includegraphics[width=\textwidth,clip=]{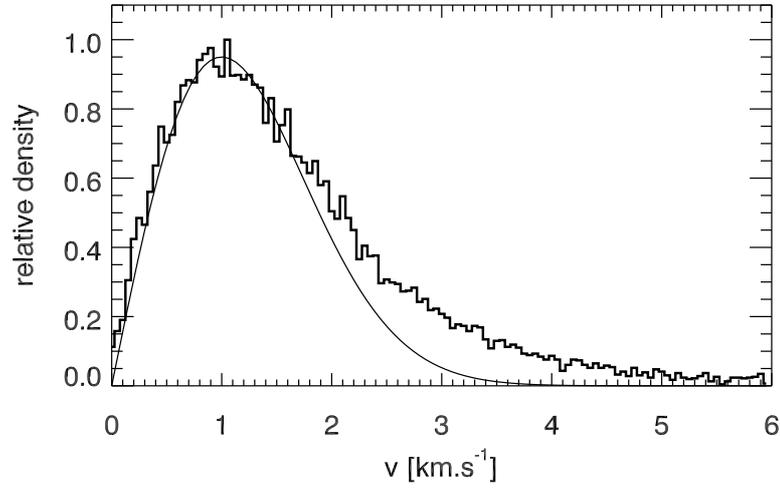}}
\caption{ An effective velocity histogram of the tracked GBPs. The number of GBPs is normalized
into the interval of 0.0-1.0 in accordance with the most probable value of the effective velocity. The thin
continuous line drawn over the histogram stands for a sample Rayleigh distribution, scaled down appropriately.}
\label{fig:2}
\end{figure}

In Figure \ref{fig:2} the distribution of effective velocities is shown together with a sample Rayleigh distribution ($\sigma$ = 1.0) scaled down to the scale of the normalized distribution of effective velocities with a good
coincidence in the interval from 0.0 to 1.5\,km\,s$^{-1}$. The obtained distribution of effective velocities compared to the sample Rayleigh distribution shows an increasing discrepancy for the velocities between $\sim$2\,km\,s$^{-1}$ and $\sim$4\,km\,s$^{-1}$.

\subsection{CHANGE IN EFFECTIVE VELOCITY} 
  \label{S-Change in effective velocity}
\par
\begin{figure} 
\centerline{\includegraphics[width=\textwidth,clip=]{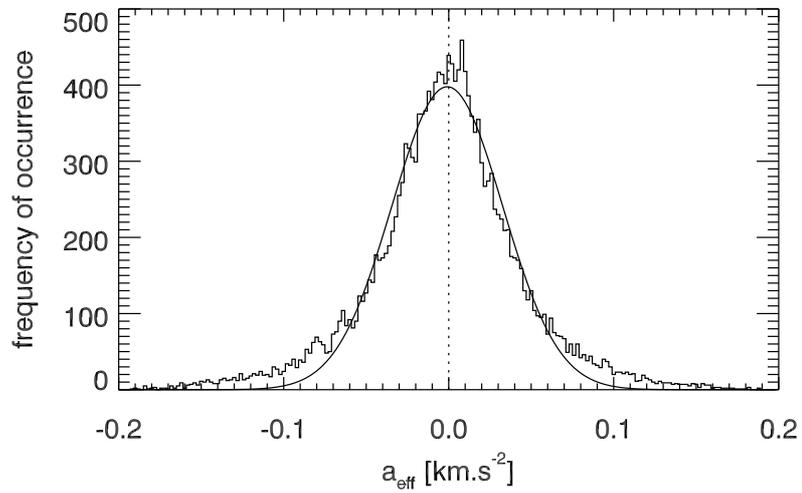}}
\caption{The $a_{\textrm{eff}}$-distribution of the tracked GBPs, fitted with a Gaussian
distribution. Positive/negative values correspond to acceleration/deceleration.}
\label{fig:3}
\end{figure}

We also calculated changes in effective velocities $a_{\textrm{\it eff}}=dv/dt$, \textit{i.e.} classical accelerations
of the tracked GBPs. Figure \ref{fig:3} shows their corresponding distribution, spanning the interval from -0.2 to
+0.2\,km\,s$^{-2}$ between two successive "identifications" of the same GBP for all studied "identifications" of GBPs. 
The obtained histogram is fitted with a Gaussian distribution (whose FWHM is 0.08\,km\,s$^{-2}$ and the
shift of the center is -0.001\,km\,s$^{-2}$). One sees that positive (acceleration) and negative (deceleration) values
occur with roughly the same probability. Most of the measured values (77.8\%) are in the range from -0.05 to +0.05\,km\,s$^{-2}$, which indicates that the effective velocity of a GBP after 30\,s increases or decreases typically by up to 1.5\,km\,s$^{-1}$ (\textit{e.g} for a GBP with the most probable velocity $\sim$0.9\,km\,s$^{-1}$, it could be an increase up to $\sim$2.4\,km\,s$^{-1}$). The excess above the Gaussian distribution seen in the wings in Figure \ref{fig:3} may be related to the excess effective velocities found between 2-4\,km\,s$^{-1}$.

\subsection{CHANGE IN DIRECTION ANGLE} 
  \label{S-Change in direction angle}
\par
To evaluate the instant direction of motion of a single GBP, we used the momentary value of the direction
angle $\varphi(t)$ defined in time step $t$ as the angle made by a given line (connecting barycenters of the brightness
of a GBP in time steps $(t)$ and $(t - 30s)$) with the axis of reference. Statistically obtained values of the
direction angle do not indicate existence of any preferred direction of motion common for all tracked
GBPs, as one would expect for such high quality data.
\par
\begin{figure} 
\centerline{\includegraphics[width=\textwidth,clip=]{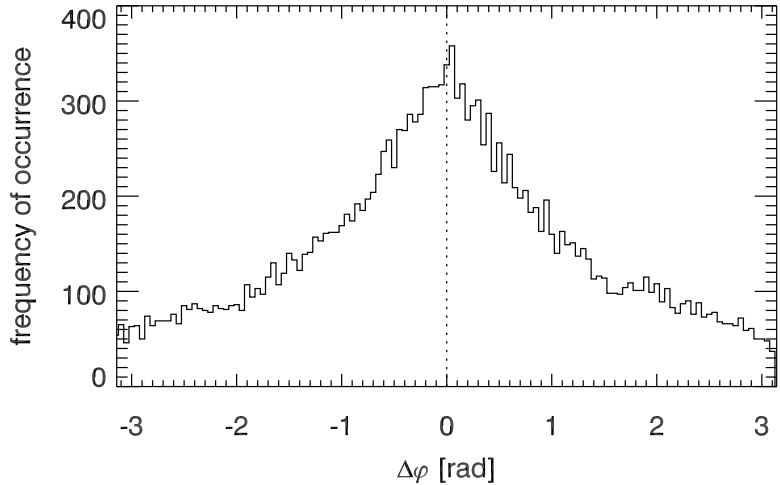}}
\caption{The $\Delta \varphi$-distribution of the tracked GBPs.}
\label{fig:4}
\end{figure}

We studied the change of the direction angle, $\Delta \varphi$, which could be used to indicate the existence of a
preferred direction of motion of a tracked GBP during its existence. This quantity is defined as $\Delta \varphi = \varphi_{2}(t_{2})-\varphi_{1}(t_{1})$, \textit{i.e.}  as the angle between directions of motion in two subsequent time steps $t_{1}$ and $t_{2}$ ($t_{2}=t_{1}+30s$) within the lifetime of a single GBP. Figure \ref{fig:4} shows a $\Delta \varphi$-distribution between two successive "identifications" of the same GBP for all studied
"identifications" of all tracked GBPs. This distribution features a broad peak, centered at $\Delta \varphi =0$, indicating
that the direction of motion between consecutive steps varies very slowly, \textit{i.e.} its change is
minimal. Nevertheless, each possible value has a nonzero probability and there even exist cases when a GBP changes direction and starts moving in the opposite direction. The distribution is a periodic one and as a whole it differs from a Gaussian distribution. The ratio of GBPs keeping their direction of motion after 30s ($-\pi/4< \Delta \varphi < +\pi/4$) to GBPs reversing their direction of motion after 30s ($\Delta \varphi <-3\pi/4$ or $\Delta \varphi >3\pi/4$) is 4.07.

\subsection{CENTRIFUGAL ACCELERATION} 
  \label{S-Centrifugal acceleration}
\par
\begin{figure} 
\centerline{\includegraphics[width=\textwidth,clip=]{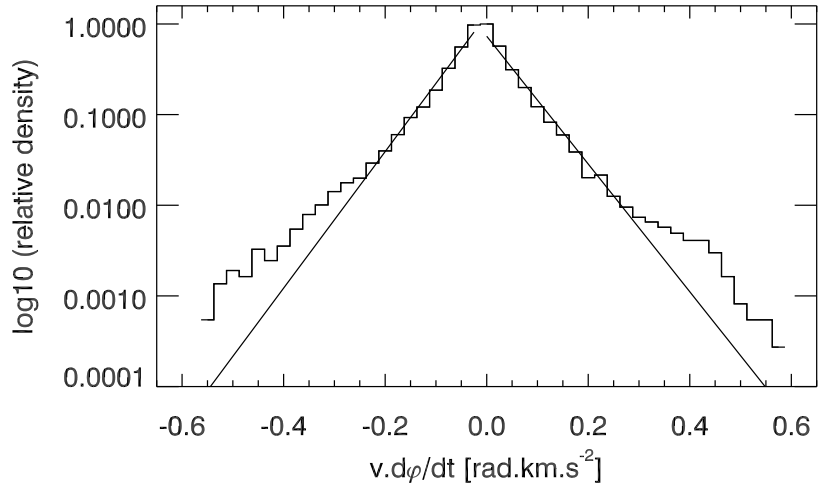}}
\caption{The distribution of centrifugal accelerations of the tracked GBPs (in a logarithmic scale). Inserted are
two linear fits, one approximating it in the range from -\bf0.25 \rm to 0.0\,rad\,km\,s$^{-2}$
(left) and the other in the range from 0.0 to 0.3\,rad\,km\,s$^{-2}$ (right).}
\label{fig:5}
\end{figure}

Centrifugal acceleration, $vd\varphi/dt$, is a relevant quantity when considering generation of transverse
waves in magnetic flux tubes \cite{2003ApJ...587..458N} and can be used to compare models of magnetic
flux tube waves with observations \cite{2006SoPh..237...13M}. Figure \ref{fig:5} shows the corresponding histogram
in a logarithmic scale. The use of logarithmic scale is motivated by a highly exponential nature of the obtained distribution, which is of a Gaussian shape (symmetric and centered around the zero value). The observed
excesses from the applied linear fits represent 0.87\% (the left-hand-side line with the slope of 5.54) and 0.42\% (the right-hand-side line with the slope of -5.48) of the measured values. 
\par
Some of the obtained values of the centrifugal acceleration exceed the value of the gravitational acceleration on the solar surface (274\,m\,s$^{-2}$ $\approx$ 0.27\,km\,s$^{-2}$), which indicates that these high values of centrifugal acceleration could have a considerable impact on the internal structure of the magnetic flux tubes (observable in the G-band as GBPs).

\subsection{RATE OF MOTION} 
  \label{S-Rate of motion}
\par
\begin{figure} 
\centerline{\includegraphics[width=\textwidth,clip=]{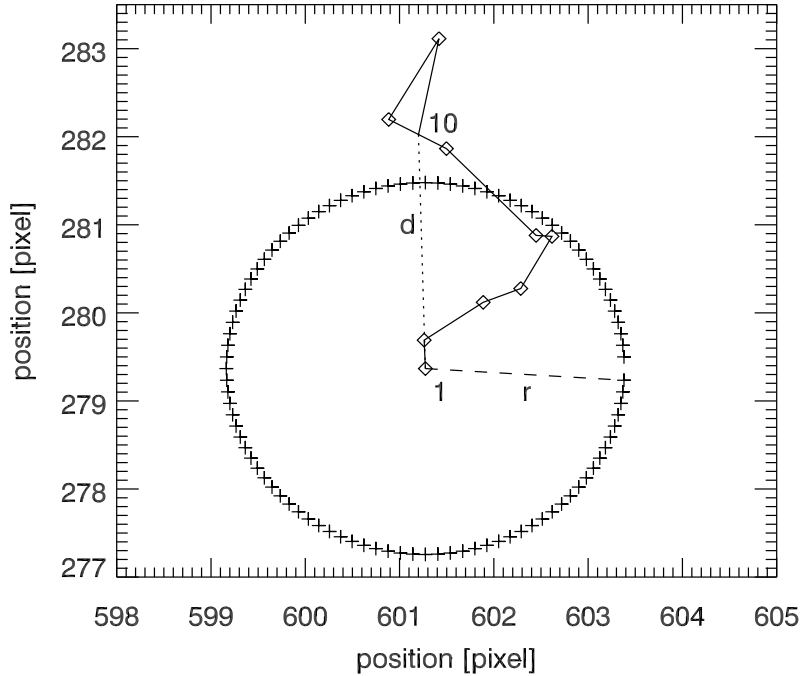}}
\caption{The track of a sample GBP. All positions except the last one are denoted by diamonds.
The dotted line represents the distance $d$ between the first and last identified position of the tracked GBP and the
dashed line stands for the radius $r$ of the circle indicating the size of the GBP at its first identification.
Numbers 1 and 10 indicate, respectively, the location of the first/last "identification" of the GBP.}
\label{fig:6}
\end{figure}

We derived the movement of numerous, randomly selected GBPs tracked from one time step to the
other throughout their existence. An example of a typical movement of a tracked GBP is shown
in Figure \ref{fig:6}. The observed movement of this GBP greatly varied in both directions and traveled
distances between consecutive measurements during its existence. Based on this example, and the other
studied GBPs, we can claim that although a typical GBP during its existence often changes its direction of
motion and makes trajectories of various lengths between subsequent time steps, the final distance
between the location (of the barycenter of brightness) of its first and last "identifications" is quite small
(mostly up to 1\,arcsec). We can thus define a new important parameter: the rate of motion $m$ of a single tracked GBP
$m=d/r$, whereas there is no correlation between $d$ and $r$. Here $d$ is the distance between the locations of the barycenters of brightness of the first and the last
"identification" of a single tracked GBP and $r$ is the radius of the circle which corresponds to the size
of the tracked GBP at its first identification. Based on the number of pixels making up the GBP on the
G-band image, the mean area of a GBP is $\sim$20\,px$^{2}$, \textit{i.e.} $\sim$0.1\,arcsec$^{2}$.
\par
\begin{figure} 
\centerline{\includegraphics[width=\textwidth,clip=]{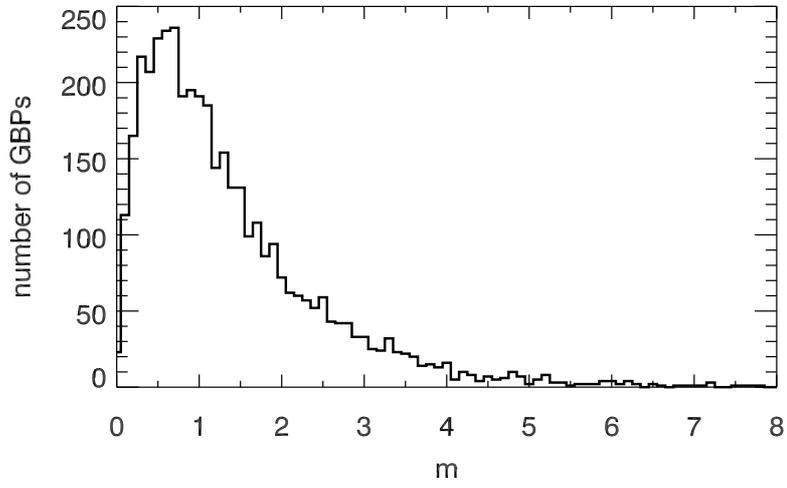}}
\caption{A histogram of the rates of motion of the tracked GBPs.}
\label{fig:7}
\end{figure}
\begin{figure} 
\centerline{\includegraphics[width=\textwidth,clip=]{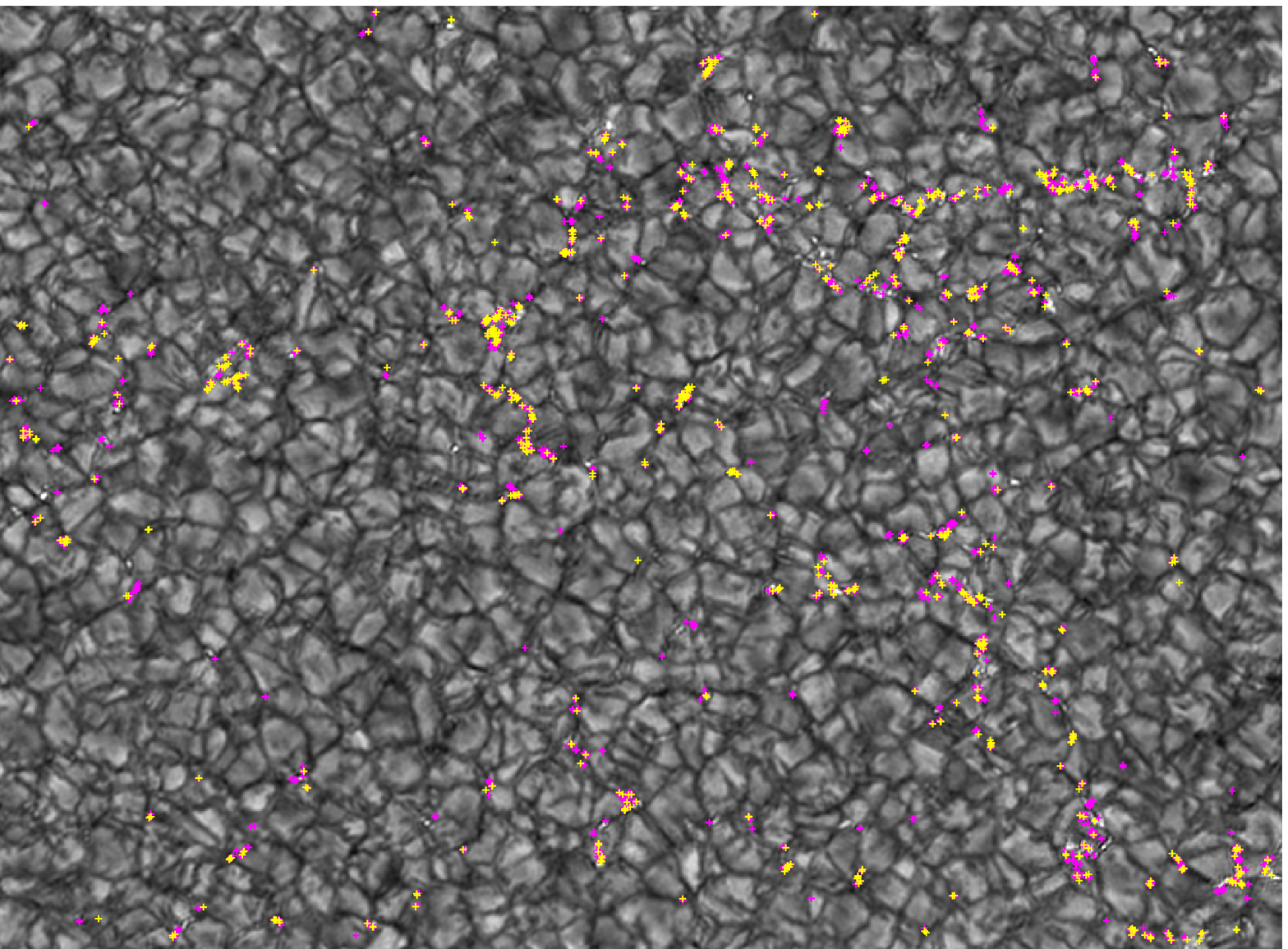}}
\caption{The full field of view of the DOT (79 $\times$ 58\,arcsec) with crosses indicating the locations of GBPs
for which the rates of motion $m < 1$ (violet) and $m > 2$ (yellow).}
\label{fig:8}
\end{figure}

Figure \ref{fig:7} shows a histogram of $m$ for all tracked GBPs. One sees that about $\sim$45\% of
them feature $m < 1$, which means that these GBPs during their existence did not move much
away from the location of their first "identification" (because their barycenters of brightness were at the
time of their last identification still inside the circle). On the other hand, those GBPs with $m > 1$ ended up their existence in the region outside the circle. Furthermore, for $\sim$18.5\,\%
of the tracked GBPs $2 < m < 4$; this means that these GBPs
show significant movement that cannot be properly accounted for by the momentary location of the
barycenter of brightness within the area of the GBP. We have been particularly interested in locations of GBPs with extreme values of the rate of motion within the field of view of the DOT. As it is implied in Figure \ref{fig:8}, the locations of nearly stable GBPs with $m < 1$
more-or-less coincide with those of farther traveling GBPs with $m > 2$, \textit{i.e.} GBPs with quite different
values of $m$ can be found in the same locations within the studied field of view. It then follows that the 
parameter $m$ is not suitable for classifying locations of GBPs.

\subsection{TIME-LAG BETWEEN RECURRENCE OF GBPS} 
  \label{S-Time-lag}
\par
As already mentioned, the tracked GBPs were preferentially located in certain places within the field
of view of the DOT. In these places, especially in the network region, a higher number of GBPs was
seen throughout the whole duration of the observation, many GBPs disappearing to be replaced by newly born ones. Therefore, we also tried to address the frequency of recurrence of different GBPs at the same
locations.
\par
\begin{figure} 
\centerline{\includegraphics[width=\textwidth,clip=]{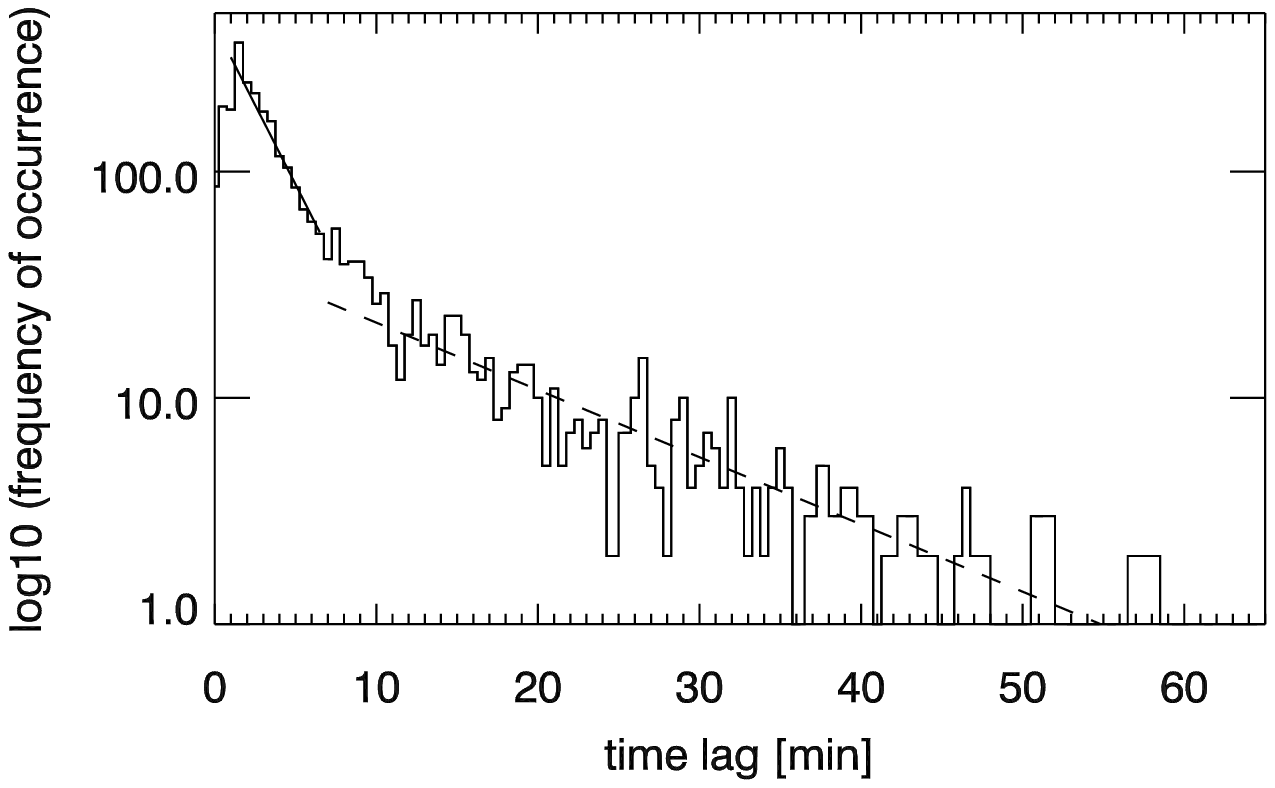}}
\caption{A histogram of time lags between reccurrence of GBPs (within a circle with the diameter of 0.36\,arcsec) in the same location within the field of view of the DOT. Inserted are two linear fits approximating
histogram values for time lags smaller (solid) and longer than 7\,min (dashed).}
\label{fig:9}
\end{figure}

We studied the area within a small circle with the radius of 0.36\,arcsec (5\,pixels) around each tracked
GBP during the whole observation in order to register occurrence of other GBPs in the same area.
Based on this study, we computed time lags between disappearance of a particular GBP and
emergence of a new one in the same area. Additionally, we repeated this process for cases with gradually increased circle's radii (0.50, 0.64, 0.78 and 0.92\,arcsec, which amounts to 7, 9, 11 and 13\,pixels).
Figure \ref{fig:9} shows the obtained histogram of time lags between recurrence of GBPs within a small circle
with the radius of 0.36\,arcsec during the whole duration of the observation in a logarithmic scale. The two
inserted lines indicate a difference in the overall trend of the obtained distribution for the values of time lag
smaller (solid line with the slope of -0.14) and higher (dashed line with the slope of -0.03) than $\sim$7 min. This difference in trends was present in the distributions of all the studied cases, which
indicates that considerably different mechanisms are responsible for time lags shorter and longer than $\sim$7\,min, respectively.
\par
The histograms obtained for circles of varying radii indicate a much higher number of small time
lags (up to $\sim$4\,min) between occurrence of GBPs at the same location. The most numerous are time lags
up to $\sim$2-3\,min. Those longer than $\sim$10\,min are much less frequent. These facts imply that the tracked
GBPs tend to disappear and reappear within a short time period, which
would indicate that the same location within the observed field of view is continuously occupied by
magnetic field concentrations, which occasionaly become visible as GBPs. These areas are sites of repetitive appearance and disappearance of numerous tracked GBPs,
while in other areas (including the extensive areas of internetwork regions) it is very rare to find a GBP during
the whole duration of the observation.

\section{Discussion}
     \label{S-Discussion}
\par
The data set used in this article was obtained by a ground-based telescope (the DOT) and the
correction of the G-band images for seeing effects was carried out through speckle reconstruction
(\opencite{2004A&A...413.1183R}; \opencite{2001ASPC..236..431S}). 
Our results can be compared to previous findings on GBPs from a variety of data sets obtained by both ground-based telescopes and space-borne instruments (SVST(SST)/La Palma - \opencite{1996ApJ...463..365B}; \opencite{1998ApJ...495..973B}; \opencite{2006SoPh..237...13M}; DOT - \opencite{2003ApJ...587..458N};
SOT/\textit{Hinode} – \opencite{2008ApJ...684.1469D}; \opencite{2009A&A...498..289U}; \citeyear{2010A&A...511A..39U}).
\par
To this end in view, it is important to take into account the resolution of our data. The typical spatial
resolution of the DOT in the G-band is $\sim$0.27\,arcsec ($\sim$200\,km) and corresponds to the reported typical size of
GBPs. The cadence of the speckle-reconstructed G-band images is 30\,s, yet GBPs could exhibit
dynamical evolution also on much smaller time scales. Since the sizes of GBPs are close to the spatial
resolution, it is difficult to get any further insight into the shapes of the tracked GBPs,
or the possible variation of their shapes during their lifetime.
\par
Also, properties of the fully automatic algorithm we employed to identify and track GBPs should be borne in mind.
The advantage of an automatic algorithm when compared to the previously used manual (\textit{e.g} \opencite{2003ApJ...587..458N}) and semiautomatic (\textit{e.g} \opencite{2006SoPh..237...13M}) ones is that it reduces the amount of subjectivity that enters the identification process. This
algorithm is similar to the so-called MLT 4 algorithm of \inlinecite{2007SoPh..243..121B}, but \-- unlike the latter that was originally developed to investigate granulation patterns
and only later adapted to investigations of GBPs \-- this one was developed to be exclusively used to investigate GBPs.
\par
The dynamics of GBPs is usually studied in terms of their effective velocity, which we chose as our
premier parameter. Its most probable value found by us ($\sim$0.9\,km\,s$^{-1}$) is a bit
smaller (not more than 30\%) than those published by \inlinecite{1998ApJ...495..973B}, \inlinecite{2003ApJ...587..458N}, \inlinecite{2006SoPh..237...13M}, and \inlinecite{2008ApJ...684.1469D}. At the same time, it lies within the range found by \inlinecite{1996ApJ...463..365B} and does not contradict those values reported by \inlinecite{2010A&A...511A..39U}. A summary of effective velocities reported by other authors is given in Table 1.
The observed scatter of these values can be attributed to the use of data obtained with
different instruments, different spatial and temporal resolution/sampling (see \textit{e.g} \opencite{2012ASPC..454...55U}), different image processing,
employment of various identification algorithms or/and by a choice of a different parameter representing
the velocity distribution. The obtained effective velocity distribution is not symmetric, and it
deviates from the sample Rayleigh function in the range of $\sim$2-4\,km\,s$^{-1}$. The observed deviation could be explained by an additional velocity supplied to GBPs by convective processes. The interaction of magnetic flux tubes (represented as GBPs) with the convective processes in the solar photosphere (dynamics of granules) may cause the distribution to include non-random values, which could be the source of the observed deviation.

 \begin{table}
 \caption{Overview of GBP velocity estimations published in various studies.}
 \label{tbl:1}
 \begin{tabular}{lc}     
 \hline
 Paper & Reported GBP velocity [km\,s$^{-1}$]\\
 \hline
 \inlinecite{1996ApJ...463..365B} & $\sim$(0.0-5)\\
 \inlinecite{1998ApJ...495..973B} & 1.47\\
 \inlinecite{2003ApJ...587..458N} & 1.31\\
 \inlinecite{2006SoPh..237...13M} & 1.11$\pm$0.7\\
 \inlinecite{2008ApJ...684.1469D} & 1.57$\pm$0.08\\
 \inlinecite{2010A&A...511A..39U} & 1-2\\
 This study & $\sim$0.9\\
 \hline
 \end{tabular}
 \end{table}

\par
The obtained histogram of changes in the direction angles of the GBP velocities is in agreement with that
published by \inlinecite{2003ApJ...587..458N} and as it is based on a bigger sample, it strengthens their claim that the direction of motion of GBPs remains the same from one time interval to the next (on the scale of 30\,s).
\par
The obtained histogram of centrifugal accelerations is also comparable to similar histograms published
by \inlinecite{2003ApJ...587..458N}, \inlinecite{2008ApJ...684.1469D} and \inlinecite{2006SoPh..237...13M}, although these authors did not use a logarithmic scale. Our distribution is, however, characterized by a richer sample of measurements. The existence of values of the centrifugal acceleration exceeding the gravitational acceleration on the solar surface (0.27\,km\,s$^{-2}$) suggests a possible change in the internal structure of the magnetic flux tube caused by the imbalance of the applied forces.    
\par
The assessment of the movement of the GBPs similarly to the previous studies (\textit{e.g} \opencite{2006SoPh..237...13M};  \opencite{2010A&A...511A..39U}), was based on the motion of their barycenters of brightness. The corresponding results can, however, be distorted because the location of the barycenter of brightness within the
area of a GBP is subject to change from one time interval to the next (caused by changes of the shape or
the brightness of the GBP). Hence, to address this issue more properly, we introduced a new
parameter called the rate of motion, which also takes into account the
size of the GBP at its first identification. As already mentioned, we found that for $\sim$45\% of the
tracked GBPs the value of their rate of motion is $m<1$, while their mean effective velocity is 0.87\,km\,s$^{-1}$. At the moment of their disappearence their barycenter positions were still inside the
circle indicating their initial size (mean size $\sim$0.1\,arcsec$^{2}$). Therefore, their displacement (\textit{i.e.} the distance between the positions of their first and last "identification") was small compared to their size.
\par
Every automatic identification and tracking algorithm has, of course, its limitations and depends also on the
temporal and spatial resolution of the used data set of G-band images. Depending on the quality of the
data and the settings of the algorithm time series of GBP "identifications" can be broken down into two
or more tracked GBPs, while they actually should be investigated as one tracked GBP; on the other hand, 
"identifications" of different cospacial GBPs can be regarded as a single tracked GBP. Taking this
possibility into account we investigated the time between recurrence of different tracked GBPs within
a selected area around the location of the barycenters of brightness of each tracked GBP. The distances
between the locations of the barycenters of brightness were selected to lie in the interval from 0.36 up to 0.92\,arcsec. We found two different trends in the distributions of time lags (all studied sizes of selected
areas) lying on both sides of the value equal $\sim$7\,min. In case of time lags smaller than $\sim$7\,min
the obtained values can be affected by the identification and tracking process (\textit{e.g} lost and found GBPs due to the tracking method). However, the values greater
than $\sim$7\,min indicate a variability of GBP occurrence caused most likely by the physical properties of the magnetic flux tubes (a change of inclination with regard to the line of sight, constriction, slanting, etc.) that are observed in G-band via their footpoints as GBPs (the possible effect of the tracking process becomes insignificant due the lenght of the time lags).
The most numerous time lags are those up to $\sim$3\,min (equal to four time steps within the duration of the time series), which indicates that certain
network regions within the field of view of the DOT are covered by GBPs during the whole duration of the
observation. While time lags of up to $\sim$4\,min are more numerous than longer time lags, there seems to
be no pronounced relation between the length of a time lag and the dominant 5-min oscillations observed in the solar photosphere.
\par
The results obtained for both new parameters (the rate of motion and the time lag) indicate that the locations of GBPs  with different characteristics (true for stable/farther traveling or GBPs with short/long lifetimes) are bound to the locations of more stable and long-lived magnetic field concentrations, which also supports the conjecture of \inlinecite{2013A&A...554A..65U} that the underlying magnetic field changes on a different (much longer) characteristic
timescale than the GBPs do. Thus, the observed disappearance/reappearance of GBPs cannot be perceived as the disappearance/reappearance of the corresponding magnetic field concentrations. This is consistent with the conclusions of \inlinecite{2001ApJ...553..449B} that the magnetic field underlying GBPs is very stable for periods of hours, and \inlinecite{2005A&A...441.1183D} that bright points trace locations of magnetic field that remain stable for long periods of time.

%

\section{Conclusions}
     \label{S-Conclusions}
\par
In this study we investigated various parameters describing dynamics of GBPs in the quiet solar
photosphere. Our results concerning effective velocity, change in direction angle and centrifugal
acceleration are acknowledging the results of previous authors, despite the use of data obtained by
different instruments, different applied image processing and the use of different identification
algorithms. It is worth to note that the data used in the presented study share comparable temporal and
spatial resolution with similar data used in other studies.
\par
The most probable value of effective velocity was $\sim$0.9\,km\,s$^{-1}$, whereas we found a deviation of the
effective velocity distribution from the expected Rayleigh function ($\sim$ sample function with $\sigma$ = 1.0) for velocities in the range
of $\sim$2-4\,km\,s$^{-1}$. The obtained change in effective velocity distribution is of Gaussian shape with FWHM
= 0.08\,km\,s$^{-2}$. On the other hand we found a non-Gaussian distribution of change in direction angle. The
obtained distribution of centrifugal acceleration is of a highly exponential nature with insignificant
excesses (0.87\% and 0.42\% of values, respectively) from a symmetric distribution.
We defined two new parameters describing dynamics of GBPs concerning the real displacement
between their appearance and disappearance (rate of motion) and the frequency of their recurrence at
the same locations (time lag between recurrence of GBPs). These new parameters can not be compared
to any parameter previously investigated in studies on GBP dynamics.
\par
We found that for $\sim$45\% of the tracked GBPs (with the values of rate of motion up to 1) their
displacement was small compared to their size. The locations of the tracked GBPs mainly include the
boundaries of supergranules representing the network, whereas there is no significant difference in the
locations of stable GBPs ($m < 1$) and the locations of GBPs with more significant displacements ($m > 2$). 
We found two different trends for the distributions of time lags for values smaller (slope of the trend line is -0.14) and greater (slope of the trend line is -0.03) than $\sim$7 min, respectively. The time lag between recurrence of GBPs shows most numerous time lags lasting $\sim$2-3\,min. Generally, time lags up to $\sim$4\,min are more numerous than longer time lags and we did not find a pronounced relation between the length of a time lag and dominant 5-min oscillations observed in the solar photosphere.
\par
Various studies indicate that the spatial resolution of modern telescopes is not able to detect the real minimum size of GBPs (see, \textit{e.g}, \opencite{2010ApJ...725L.101A}). Many of the GBPs identified and tracked for our study may consist of clusters of yet unresolved GBPs. Moreover, our study of dynamics suggests that the movement of these small-scale magnetic elements occurs on much smaller timescales (shorter than 30\,s) and within a much smaller space than our data allow us to investigate. Thus for a more detailed study of their trajectories data sets with higher temporal and spatial resolution are needed.
\par
With our new parameters (rate of motion and time lag between recurrence of GBPs) we also call
attention to additional uncertainties in evaluation of dynamics of GBPs caused by the settings of the
applied identification and tracking algorithm as much as by the low time and spatial resolution of the
studied data set.

%

%
 \begin{acks}
The Technology Foundation STW in the Netherlands financially supported the
development and construction of the DOT and follow-up technical developments. The DOT has been built by instrumentation groups of Utrecht University and Delft University (DEMO) and several firms with specialized tasks. The DOT is located at Observatorio del Roque de los Muchachos (ORM) of Instituto de Astrof\'{\i}sica de Canarias (IAC). DOT observations on 19 October 2005 have been funded by the OPTICON Trans-national Access Programme and by the ESMN-European Solar Magnetic
Network - both programs of the EU FP6. The authors thank P. S\"utterlin for the DOT observations and R. Rutten for the data reduction. This work was supported by the Slovak Research and Development Agency under the contract No. APVV-0816-11.We acknowledge support by the project VEGA 2/0108/12. We are indebted to an anonymous reviewer whose comments improved the paper substantially.
 \end{acks}

%
%
\bibliographystyle{spr-mp-sola}
%

\end{article} 
\end{document}